\documentclass[grl]{agutex}

\usepackage{amssymb,amsthm,graphics,graphicx,epsfig,amsmath}
\authorrunninghead{N\'U\~NEZ VALDEZ ET AL.}

\titlerunninghead{Thermoelasticity of $\alpha-\beta-$(Fe$_x$,Mg$_{1-x}$)$_2$SiO$_4$}

\authoraddr{M. N\'u\~nez-Valdez,
School of Physics and Astronomy, University of Minnesota, Minneapolis, Minnesota, USA. (valdez@physics.umn.edu)}

\authoraddr{Z. Wu,
School of Earth and Space Sciences, University of Science and Technology of China, Hefei, Anhui 230026, China.
(wuzq10@ustc.edu.cn)}

\authoraddr{Y. G. Yu,
Department of Chemical Engineering and Materials Science, University of Minnesota, Minneapolis, MN 55414, USA.
(yuxxx135@umn.edu)}

\authoraddr{R. M. Wentzcovitch, 
Department of Chemical Engineering and Materials Science, Minnesota Supercomputing Institute (MSI), University of Minnesota, Minneapolis, Minnesota, USA.(wentzcov@cems.umn.edu)}

\begin{document}

%
%

\title{Thermoelastic Properties of Olivine and Wadsleyite (Fe$_x$,Mg$_{1-x}$)$_2$SiO$_4$: Their Relationship to the 410 km Seismic Discontinuity.}
%
%

%
%









%
%


\begin{abstract}
We combine density functional theory (DFT) within the local density approximation
(LDA), the quasiharmonic approximation (QHA), and a model
of vibrational density of states (VDoS) to calculate elastic moduli and sound
velocities of $\alpha-$ and $\beta-$(Fe$_x$,Mg$_{1-x}$)$_2$SiO$_4$ (olivine and wadsleyite), the most abundant minerals of the Earth's upper mantle (UM) and upper transition zone (TZ). Comparison with experimental values at room-temperature
and high pressure or ambient-pressure and high temperature show good
agreement with our first-principles findings. Using our results, we investigate
the discontinuities in elastic moduli and velocities associated with the
$\alpha\rightarrow\beta-$(Fe$_x$,Mg$_{1-x}$)$_2$SiO$_4$ transformation at pressures and temperatures relevant to the 410 km seismic discontinuity. We find the compressional velocity contrast  to be smaller than the shear velocity contrast, in agreement with the preliminary reference earth model (PREM).
\end{abstract}

%
%

%

\begin{article}

%
%

\section{Introduction}
Wadsleyite ($\beta-$phase) is a high-pressure polymorph of olivine ($\alpha-$phase), (Mg$_{1-x}$,Fe$_x$)$_{2}$SiO$_{4}$. These minerals are the main constituents of the Earth's upper mantle (UM) and upper transition zone (TZ) \citep{Rinwood,IR,Putnis}. At $\sim$13.5 GPa, the transformation from olivine to wadsleyite happens near 1600 K \citep{Ak89,KI89}. This transformation is associated with the 410 km discontinuity in seismic velocities in the Earth. The interpretation of seismic models in terms of mineralogy and chemical composition requires knowledge of  bulk ($K$) and shear ($G$) moduli as functions of pressure and temperature of constituent minerals. 
Elastic properties of iron-bearing olivine have been studied experimentally at ambient temperature using impulsively stimulated laser scattering (ISLS) to 17 GPa \citep{Abramson}, Brillouin spectroscopy (BS) to 32 GPa \citep{Zha98}, ultrasonic interferometry (UI) at elevated pressure and temperature \citep[see e.g.][]{Liu05}, and resonant ultrasound spectroscopy (RUS) to 1500 K \citep{Isaak92}. Elastic measurements on iron-bearing wadsleyite have been more limited in temperature and pressure range than for olivine. Data have been obtained using UI at high pressure and high temperature \citep{Li00,Liu09}, RUS and resonant sphere techniques at ambient pressure and high temperature \citep{Maya04,Isaak10}. 

Computational predictions of iron-bearing $\alpha-$ and $\beta-$phases have been reported only for static elasticity at high pressure \citep{Nunez10,Nunez11,Stack10}. High temperature first-principles calculations employing the quasiharmonic approximation (QHA), valid up to $\sim2/3$ of the melting temperature, or molecular dynamics (MD) methods, fundamental near and above melting temperatures, complement each other and have been used to obtain elastic moduli. These calculations are, however, restricted in number since they are computationally demanding \citep{Karki99,Wentz04}. QHA calculations required vibrational density of states (VDoS) for strained atomic configurations at several pressures, that is, about 1000 parallel jobs \citep{DS07,DS08,DS}. In this paper we use a new analytical and computational procedure \citep{Wu11} to calculate $K$, $G$ and compressional ($V_P$), shear ($V_S$), and bulk ($V_\Phi$) velocities of $\alpha-\beta-$(Mg$_{1-x}$,Fe$_x$)$_{2}$SiO$_{4}$ phases. So far this method has only been tested on MgO and $\alpha-$Mg$_2$SiO$_4$ at room pressure and high temperature. This approach uses only static elastic constants and phonon density of states for unstrained configurations, therefore reducing the amount of computational time, resources, and especially human effort,  by one to two orders of magnitude. We then address contrasts across the  $\alpha\rightarrow\beta$ (Fe$_x$,Mg$_{1-x}$)$_2$SiO$_4$ transition near conditions of the 410 km seismic discontinuity.  
\begin{figure*}
\begin{center}
\includegraphics[width=169mm,height=90mm]{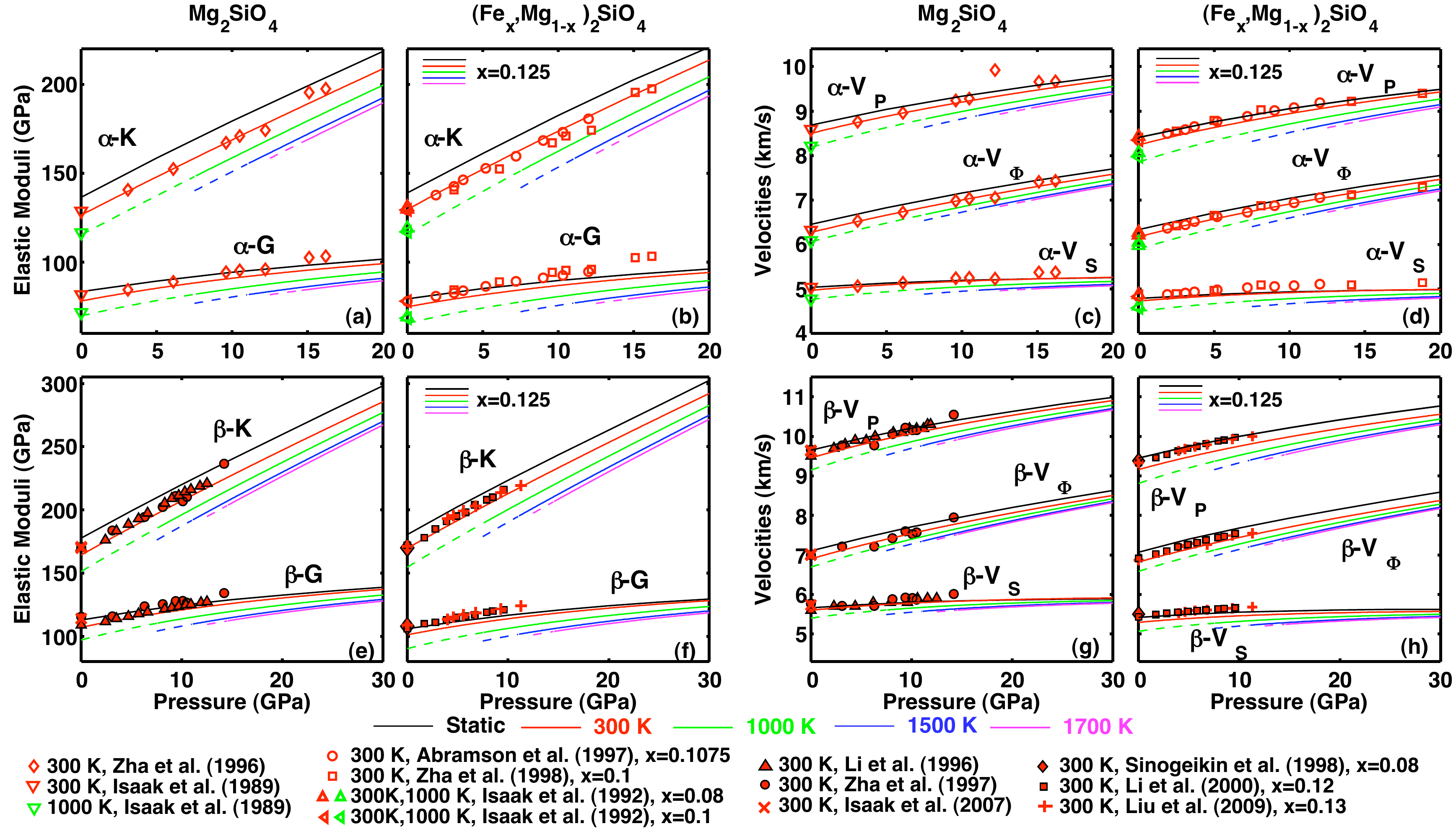}
\caption{(Color online) Pressure and temperature dependence of elastic moduli $K$ and $G$ (calculated as Voigt-Reuss-Hill averages \protect\citep{Watt}),  and velocities $V_P$, $V_S$, and $V_\Phi$ for Fe-free olivine-wadsleyite (a, c, e, g) and Fe-bearing olivine-wadsleyite (b, d, f, h). First principles calculations within LDA (lines) are compared to available experimental data (symbols). Low-pressure-high temperature trends (dash lines) are outside the validity of the QHA.}\label{Fig1}
\end{center}
\end{figure*}
\section{Theory}
The computational approach is based on density functional theory (DFT) \citep {HK,KS} within the local density approximation (LDA) \citep{Ceperley}. At arbitrary pressures ($P$'s), orthorhombic equilibrium structures  of olivine (space group Pbnm, 28 atoms/unit cell)  and wadsleyite (space group Imma, 28 atoms/primitive cell) were found using pseudopotentials, energy cutoffs, and {\bf k}-point samplings as reported by  \cite{Nunez10,Nunez11}. Interatomic forces and stresses were smaller than $10^{-4}$ Ry/a.u. Dynamical matrices were calculated using density functional perturbation theory (DFPT) \citep{Baroni2001} via quantum-ESPRESSO \citep{QE}. At each pressure, a dynamical matrix was obtained on a $2\times2\times2$ {\bf q}-point mesh for one atomic configuration only. In principle about 10 other configurations should be used as well, but here we are interested in elastic, not thermodynamic properties, and on the dependence of averages of phonon frequencies with strains, thus the current approximation is sufficiently accurate \citep{Wu11,Nunez10}. Force constants were extracted and interpolated on a $12\times12\times12$ regular {\bf q}-point mesh to produce VDoS. 

Then we exploit the dependence of phonon frequencies on anisotropic composition to determine the thermal contribution to the Helmholtz free energy $F$, within the QHA \citep{Wallace}, that is,
\begin{eqnarray}
F\left(e,V,T\right)=U_{st}(e,V)+\frac{1}{2}\sum_{{\bf q},m}{\hbar\omega_{{\bf q},m}(e,V)}+\nonumber\\
+k_BT\sum_{{\bf q},m}\ln\left\{1-exp\left[-\frac{\hbar\omega_{{\bf q},m}(e,V)}{k_BT}\right]\right\},
\end{eqnarray}
where ${\bf q}$ is the phonon wave vector, $m$ is the normal mode index,  $T$ is temperature, $U_{st}$ is the static internal energy at equilibrium volume $V$ under isotropic pressure $P$ and infinitesimal strain $e$, $\hbar$ and $k_B$ are Planck and Boltzmann constants, respectively. Isothermal elastic constants are given by $C_{ijkl}^T=\left[\frac{\partial^2G(P,T)}{\partial e_{ij}\partial e_{kl}}\right]_{P,T}$ with $G=F+PV$, the Gibbs energy and $i,j,k,l=1,\dots,3$. To convert to adiabatic elastic constants, we use $C_{ijkl}^S=C_{ijkl}^T+\frac{T}{VC_V}\frac{\partial S}{\partial e_{ij}}\frac{\partial S}{\partial e_{kl}}\delta_{ij}\delta_{kl}$, ($C_V=$heat capacity at constant $V$, and $S=$entropy).  $C_{ijkl}^T$ can be expressed as functions of strain and mode Gr\"unesein parameters. Assuming that the angular distribution of phonons is isotropic, isothermal elastic constants can be calculated without performing phonon calculations for strained configurations \citep{Wu11}. This approximation is equivalent to assuming that thermal pressure is isotropic. Even though this assumption is not completely accurate \citep{Carrier2007}, the method is a good approximation within experimental uncertainties \citep{Carrier08}. Static elastic constants previously computed \citep{Nunez10,Nunez11} are also used.
\section{Results and Discussion}
To the best of our knowledge, this is the first DFT-based report of aggregate properties of Fe-bearing olivine and wadsleyite at pressures and temperatures relevant to the UM and TZ.  Fig. (\ref{Fig1}) shows the calculated pressure dependence of $K$, $G$, $V_P=\sqrt{(K+\frac{4}{3}G)/\rho}$, $V_S=\sqrt{G/\rho}$, and $V_\Phi=\sqrt{K/\rho}$, with $\rho=$density, of olivine and wadsleyite at several temperatures compared to experimental data. The general agreement between our predictions and measurements at room temperature and 1000K  is excellent for $x=0$  \citep{Zha96, Isaak89,Li96,Zha97,Isaak07} within the valid regime of the QHA \citep{Yu08}. For $x=0.125$ our trends compare very well with experiments in the range $x=0.08-0.013$ \citep{Abramson,Zha98,Isaak92,Sinog98,Li00,Liu09}. $K$ and $G$ increase with pressure, but the pressure dependence of $K$ is stronger. Temperature decreases both $K$ and $G$ but the former is the most affected, this effect propagates to $V_P$ and $V_\Phi$ for both $x=0$ and $x=0.125$, Fig. (\ref{Fig1}).
\begin{figure*}
\begin{center}
\includegraphics[width=169mm]{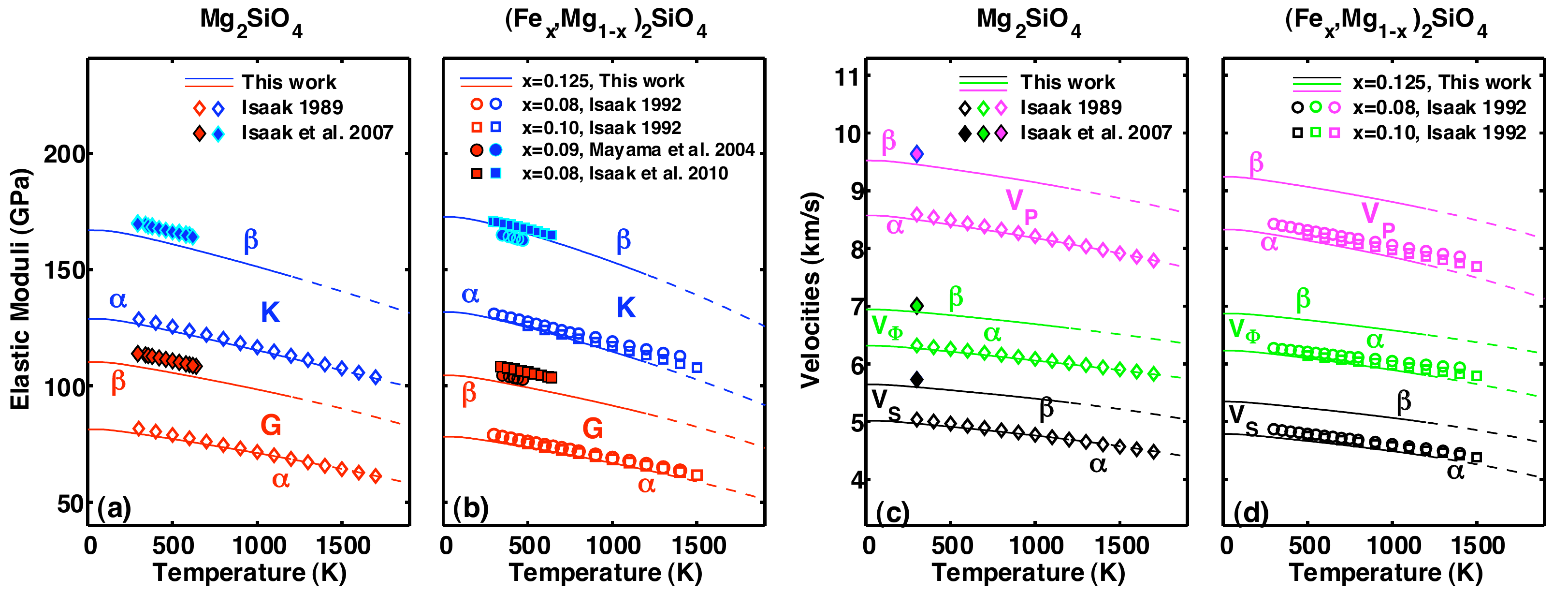}
\caption{(Color online) Temperature dependence of $K$, $G$, $V_P$, $V_S$, and $V_\Phi$ for Mg-end members olivine-wadsleyite (a, c) and Fe-bearing olivine-wadsleyite (b, d). First-principles calculations within LDA (lines) are compared to available experimental data (symbols) at P=0 GPa.}\label{Fig2}
\end{center}
\end{figure*}

Fig. (\ref{Fig2}) displays aggregate properties of $\alpha-$ and $\beta-$phases as functions of temperature at ambient pressure. The prediction power of our DFT-based calculations is outstanding within the limit of the QHA for Mg-end member and Fe-bearing  $\alpha-$phases, [see Figs. (\ref{Fig2}a and c)], even though the iron content in experimental data is somewhat different from $x=0.125$ \citep{Isaak89,Isaak92}. 
\begin{figure}
\begin{center}
\includegraphics[width=80mm]{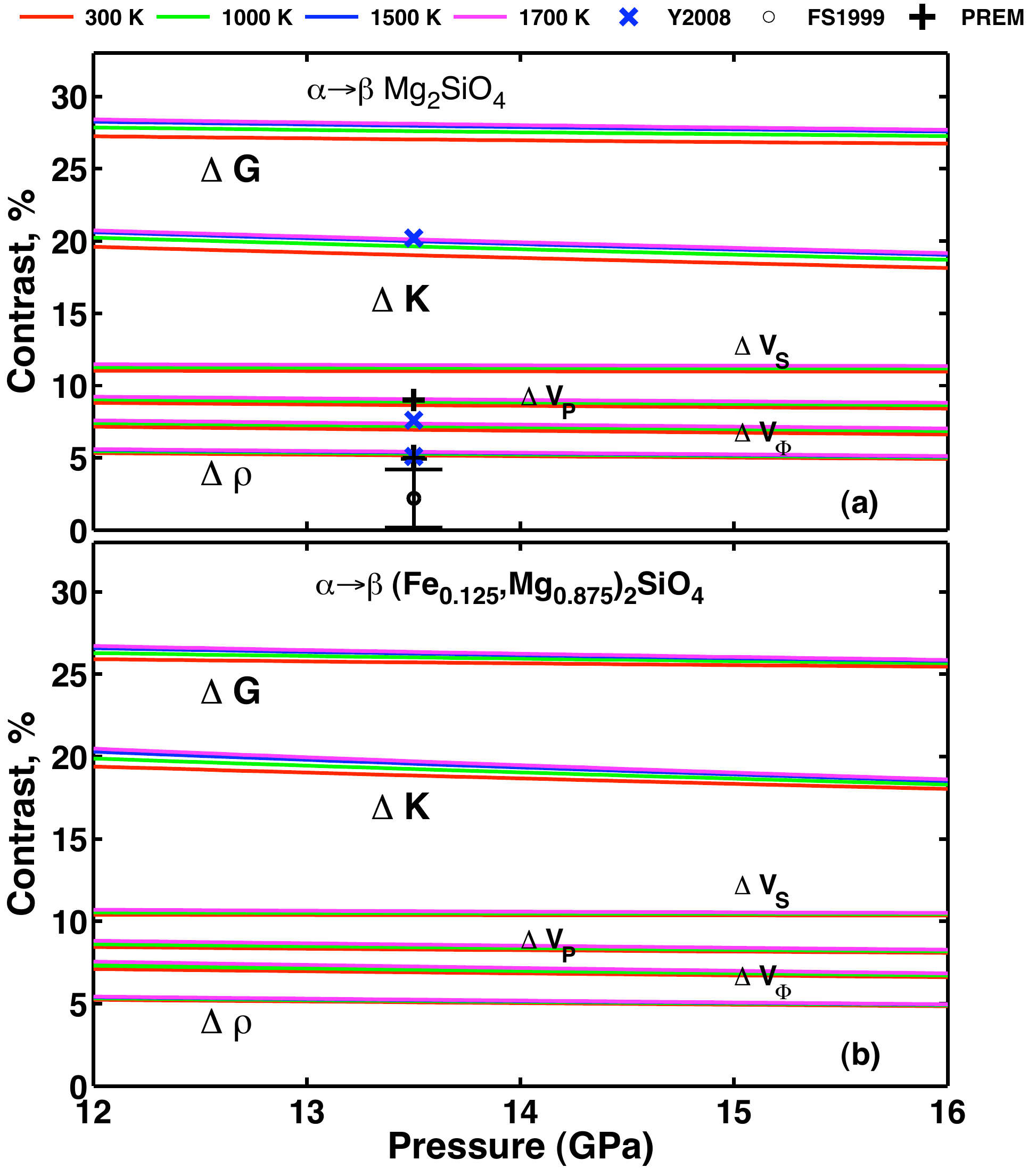}
\caption{(Color online) Density, elastic, and velocity contrasts (lines) compared to laboratory and seismic data across the Fe-free (a) and Fe-bearing (b) $\alpha\rightarrow\beta$ transition. Y2008 \protect\cite{Yu08}, SF \protect\cite{SF99}, PREM \protect\cite{PREM}.}\label{Fig3}
\end{center}
\end{figure}
At ambient conditions, our results for $K$ and $G$ agree quite well with experimental measurements for $\alpha-$ and $\beta-$Mg$_2$SiO$_{4}$ phases \citep{ThisWork}. We find that iron content increases $K$ and decreases $G$ for both phases. Experimentally this effect is small in the case of olivine, Figs. (\ref{Fig2}a) and (\ref{Fig2}b). For wadsleyite, measurements show that $K$ is even more  independent of Fe content within experimental uncertainty, while $G$ seems to decrease, in agreement with our findings. Furthermore, looking at the temperature dependence of $\beta$-$K$ at ambient pressure [Fig. (\ref{Fig2}b)] we see that our predicted trend for $x=0.125$ falls in between two experimental sets of data for $x=0.08$ and $x=0.09$, with measurements for $x=0.09$ \citep{Maya04} being smaller  than those for $x=0.08$ \citep{Isaak10}. These observations show that it is uncertain from experiments whether iron in the range $x=0.00-0.10$ causes $K$ to increase or decrease since its effect on $K$ is so small.  In contrast, calculations show clearly the dependence of $K$ on $x$.

Acoustic velocities of $\alpha-$ and $\beta-$phases decrease with increasing $x$ at all pressures and temperatures, [see Fig. (\ref{Fig1}c,d,g, and h) and Fig. (\ref{Fig2}c and d). Our results for both phases with $x=0$ are in very good agreement with experimental values, Fig. (\ref{Fig1}c and g). For $x=0.125$, our velocities are smaller than measurements in samples with $x$ in the range $0.08-0.12$, Fig. (\ref{Fig1}d and h). 
Accurate data on elasticity of $\alpha-$ and $\beta-$Mg$_2$SiO$_{4}$ at  UM and TZ conditions are critical for investigating the role of the $\alpha-$ to $\beta$ transformation on the 410 km seismic discontinuity. Predicted contrasts, $\Delta$, for a property $M$ across a $\alpha\rightarrow\beta$ transformation, $\Delta M=2\left(M_{x,\beta}- M_{x,\alpha}\right)/\left(M_{x,\alpha}+ M_{x,\beta}\right)\times 100$, indicate major changes in elastic moduli and velocities in these two phases near 410 km depth conditions, Table (\ref{table_oliv_contrasts}). Fig. (\ref{Fig3}) shows how contrasts slightly decrease with pressure and increase with temperature. The former result has been previously inferred for $x=0$ \citep{Zha98,Sinog98}, we now show that iron content does not change this trend.  We also find that iron in the aggregate does not alter the density discontinuity if its presence does not change the volume ratio between phases as it was speculated by \cite{Yu08}. The 0.2-4\% density increase estimated by a seismic impedance study \citep{SF99} is consistent with the discontinuity produced in a pyrolite-type aggregate, as discussed by \cite{Yu08}. Overall, iron decreases only slightly magnitudes of contrasts with respect to those in the Mg-end member. From our calculations we observe that  $\Delta K <\Delta G$ and $\Delta V_P<\Delta V_S$, which is consistent with contrasts from PREM \citep{PREM}.
\begin{table}
\centering          
\begin{tabular}{ l | r | r | r | r}
\hline\hline
T (K) &\multicolumn{2}{c |} {1500} &\multicolumn{2}{c }{1700 }\\
\hline
$x$& 0.0&0.125&0.0&0.125\\
\hline
$\Delta \rho$ & 5.39 & 5.22 & 5.43 & 5.25\\
$\Delta K$ & 19.99 & 19.56 & 20.12 & 19.71\\
$\Delta G$ & 27.97  & 26.25 & 28.12 & 26.36\\
$\Delta V_P$ & 9.02 & 8.54 & 9.07 & 8.59\\
$\Delta V_S$ & 11.37 & 10.58 & 11.42 & 10.62\\
$\Delta V_\Phi$ & 7.33 & 7.20 & 7.38 & 7.26\\
$\Delta (\rho V_P)$ & 14.39 & 14.23 & 14.48 & 14.30 \\
$\Delta (\rho V_S)$ & 16.73 & 16.57 & 16.82  & 16.65\\
\hline\hline
\end{tabular}
\caption{Predicted contrasts in \% across the $\alpha\rightarrow\beta$ transition at 13.5 GPa.}\label{table_oliv_contrasts}
\end{table}
\section{Conclusions}
For the first time, we have presented parameter-free first-principles results of high pressure-temperature aggregate elastic properties and sound velocities of Fe-bearing phases of olivine and wadsleyite. We used the QHA and a novel method of calculating elasticity at high temperatures without obtaining VDoS of strained configurations \citep{Wu11}. Treatment of strain Gr\"uneisen parameters via isotropic averages reduced greatly the computational cost of the task, which, otherwise, would have been prohibitively lengthy. Experiments dealing with simultaneous high pressures and temperatures offer limited data and, often relying on ambient conditions, either in temperature or pressure, to extrapolate to conditions near 410 km ($\sim$13.5 GPa and 1500 K). Here, we have shown well defined changes in elastic and acoustic properties of Fe-bearing $\alpha-$ and $\beta-$phases. Contrasts across the $\alpha$ to $\beta$ transition are clearly specified near conditions of the 410 km seismic discontinuity. Similar sets of results for other coexisting phases in the UM and TZ such as pyroxenes, garnets and Ca-perovskite are necessary. The consideration of the effects of water incorporation in both olivine and wadsleyite at relevant conditions of the 410 km discontinuity are also critical for this investigation.

%
%
%
%
%
%
%

\begin{acknowledgments}
Research supported by the NSF/EAR-1019853, and EAR-0810272.
Computations were performed using the VLab cyberinfrastructure at the Minnesota Supercomputing Institute.
\end{acknowledgments}

\end{article}


%
%

%
%
%
%
%
%
%


\end{document}